\documentclass{aa} 

\usepackage{graphicx}
\usepackage{txfonts}
\usepackage[colorlinks=true, linkcolor=blue, citecolor=blue, urlcolor=gray]{hyperref}
\usepackage{xcolor}
\usepackage{listings}
\usepackage{braket}
\usepackage{physics}

\newcommand{\txd}{{\text{d}}}

\newcommand{\bfx}{{\boldsymbol{x}}}
\newcommand{\bfn}{{\boldsymbol{n}}}
\begin{document}

   \title{Self-consistent dust and non-LTE line radiative transfer with SKIRT}


   \author{Kosei Matsumoto
          \inst{1, 2, 3}
          \and
          Peter Camps\inst{1}
          \and
          Maarten Baes\inst{1}
          \and
          Frederik De Ceuster\inst{4}
          \and
          Keiichi Wada\inst{5, 6, 7}
          \and
          \\
          Takao Nakagawa\inst{2, 3}
          \and
          Kentaro Nagamine\inst{8, 9, 10}
          }

   \institute{
            Sterrenkundig Observatorium, Universiteit Gent, Krijgslaan 281 S9, B-9000 Gent, Belgium, \\
            \email{kosei.matsumoto@ugent.be}
            \and
            Department of Physics, Graduate School of Science, The University of Tokyo, 7-3-1 Hongo, Bunkyo-ku, Tokyo 113-0033, Japan
            \and
            Institute of Space and Astronautical Science, Japan Aerospace Exploration Agency, 3-1-1 Yoshinodai, Chuo-ku, Sagamihara, Kanagawa 252-5210, Japan
            \and
            Institute of Astronomy, KU Leuven, Celestijnenlaan 200D, 3001 Leuven, Belgium
            \and
            Kagoshima University, Graduate School of Science and Engineering, Kagoshima 890-0065, Japan
            \and
            Ehime University, Research Center for Space and Cosmic Evolution, Matsuyama 790-8577, Japan
            \and
            Hokkaido University, Faculty of Science, Sapporo 060-0810, Japan
            \and
             Theoretical Astrophysics, Department of Earth and Space Science, Osaka University, 1-1 Machikaneyama, Toyonaka, Osaka 560-0043, Japan
            \and
            Kavli IPMU (WPI), The University of Tokyo, 5-1-5 Kashiwanoha, Kashiwa, Chiba 277-8583, Japan
            \and
            Department of Physics and Astronomy, University of Nevada, Las Vegas, 4505 S. Maryland Pkwy, Las Vegas, NV 89154-4002, USA
          }

   \date{\today}

 
\abstract{%
We introduce Monte Carlo-based non-LTE line radiative transfer calculations in the 3D dust radiative transfer code {\sc SKIRT}, which was originally set up as a dust radiative transfer code. By doing so, we develop a generic and powerful 3D radiative transfer code that can self-consistently generate spectra with molecular and atomic lines against the underlying continuum. We test the accuracy of the non-LTE line radiative transfer module in the extended {\sc SKIRT} code using standard benchmarks. We find excellent agreement between the {\sc SKIRT} results, the published benchmark results, and results obtained using the ray-tracing non-LTE line radiative transfer code {\sc MAGRITTE}, which validates our implementation. We apply the extended {\sc SKIRT} code on a 3D hydrodynamic simulation of a dusty AGN torus model and generate multi-wavelength images with CO rotational-line spectra against the underlying dust continuum. We find that the low-$J$ CO emission traces the geometrically thick molecular torus, whereas the higher-$J$ CO lines originate from the gas with high kinetic temperature located in the innermost regions of the torus. Comparing the calculations with and without dust radiative transfer, we find that higher-$J$ CO lines are slightly attenuated by the surrounding cold dust when seen edge-on. This shows that atomic and molecular lines can experience attenuation, an effect that is particularly important for transitions at mid- and near-infrared wavelengths. Therefore, our self-consistent dust and non-LTE line radiative transfer calculations can help interpret the observational data from \textit{Herschel}, \textit{ALMA}, and \textit{JWST}.
}

   {}


   \maketitle
%

\section{Introduction}\label{introduction}

Continuum and spectroscopic observations can reveal a wealth of information on the dusty and gaseous medium for many astrophysical objects, such as the atmospheres of stars and exoplanets, protoplanetary disks and star-forming regions, and the interstellar medium (ISM) of galaxies. Continuum observations at infrared and submillimetre wavelengths can be used to determine the mass and temperature distribution of the dust in the ISM \citep{Galliano2018}. Doppler shifts of molecular and atomic lines provide powerful clues to understanding gas kinematics \citep{Bland-Hawthorn2016}, and the line ratio provides us with thermal conditions and chemical composition in the gas \citep[][]{Wolfire2022}. 

Today, state-of-the-art facilities such as the James Webb Space Telescope (\textit{JWST}) and the Atacama Large Millimeter/submillimeter Array (\textit{ALMA}) provide us with high-quality spectroscopic and photometric data, which reveal various astrophysical phenomena on the dust, atoms, and molecules within the ISM of galaxies \citep[e.g.,][]{Alvarez-Marquez2022, Rich2023, Liu2023, Meidt2023, Appleton2023, Schinnerer2023, Leroy2023}. For the interpretation of these observational data, a basic approach is a comparison between observations and three-dimensional (3D) hydrodynamic simulations, which predict the spatial distribution, velocity structure, and physical conditions of the multi-phase ISM in galaxies \citep[e.g.,][]{2017MNRAS.466.1903G, 2018ApJ...853..173K, 2020ApJ...897..143K, 2021MNRAS.504.1039R, 2023ApJ...946....3K}, although inferring the physical properties of the ISM even from these high-quality observations is not trivial. Observations are always projections along the line of sight, and detailed information about the complex 3D structure of the objects under study is lost. Moreover, differential dust attenuation along the line of sight affects the relative strengths of the lines and complicates the interpretation. In addition, the observational analysis methods used to extract physical information from the observed properties invariably involve assumptions and systematic uncertainties (e.g., calibrations of the continuum subtraction or assumption of local thermodynamic equilibrium (LTE) for the excitation states in molecules and atoms). Therefore, forward modeling synthetic observational data from hydrodynamic simulations is an important means to help us understand the observational data and assess the importance of assumptions and systematics. 


For an accurate estimation of the line intensities from atoms and molecules, we need non-LTE line radiative transfer codes that self-consistently calculate, at every location in the simulation volume, the radiation field and the level populations considering collisional and radiative transitions due to external radiation. Such non-LTE calculations are essential because the ISM is composed of not only dense gas clouds in LTE conditions but also diffuse gas clouds exposed to radiation. Over the past years, many non-LTE line radiative transfer codes, most of them ray-tracing codes, have been developed, including {\sc RATRAN} \citep{Hogerheijde&vanderTak2000}, {\sc LIME} \citep{Brinch2010}, {\sc TORUS} \citep{Rundle2010}, {\sc MOLLIE} \citep{2010ApJ...716.1315K}, {\sc 3D-PDR} \citep{Bisbas2012}, {\sc LOC} \citep{Juvela2020}, and {\sc MAGRITTE} \citep{Frederik2020}.  
To calculate radiation fields, these codes backtrack the paths of incident rays on each point and average the numerically calculated intensities of the incoming rays.
While these codes accurately calculate radiation fields when cosmic microwave background (CMB) is dominant as external radiation, it is difficult to handle dust radiative processes including scattering processes.
As a result, their calculations have never involved dust radiative transfer, and are not good at estimating spectra of atomic and molecular lines in the near- and mid-infrared (NIR and MIR, respectively) wavelength ranges, where dust emission is dominant.

On the other hand, to generate mock continuum infrared dust observations, dust radiative transfer codes are required. Typically, they are stochastic Monte Carlo codes, simulate the emission of heating sources and the absorption, scattering, and thermal emission by the dust using a finite number of photon packets, and calculate radiation fields by averaging the intensities of the photon packets passing through each cell. 
In this way, the temperature distribution of the dust grains and corresponding emission are estimated at every location in the simulation volume.
3D dust radiative transfer has experienced a strong development over the past two decades \citep{2013ARA&A..51...63S}, and nearly all modern 3D dust radiative transfer codes, such as {\sc DIRTY} \citep{2001ApJ...551..269G}, {\sc SKIRT} \citep{Baes2011}, {\sc HYPERION} \citep{Robitaille2011}, {\sc RADMC-3D} \citep{Dullemond2012}, {\sc POLARIS} \citep{2016A&A...593A..87R}, and {\sc SOC} \citep{2019A&A...622A..79J}, employ the stochastic Monte Carlo method. 

Ideally, dust and non-LTE line radiative transfer postprocessing should be done with a single code. This has the obvious advantage that the calculations can be done simultaneously and self-consistently.
An accurate estimation of the dust emission is essential for atomic and molecular lines against the dust continuum since the dust emission can excite them \citep{Matsumoto2022}. In fact, some observations report that IR radiation pumps the rotational and vibrational states of molecules \citep[e.g.,][]{Aalto2007, Imanishi2016, Imanishi2017, Runco2020}. Moreover, dust radiative transfer is also needed to calculate the attenuation of the line emission itself, in particular for NIR and MIR lines such as the OH and H$_2$ rotational lines and CO rovibrational lines. To simultaneously perform non-LTE line and dust continuum radiative transfer, the Monte Carlo approach seems to be the most suitable option. The technique is conceptually straightforward and intrinsically 3D, which makes it naturally suitable for complex 3D structures. Moreover, scattering is, contrary to ray-tracing methods, easily incorporated in the Monte Carlo framework \citep{2013ARA&A..51...63S}. 

There are some challenges to the establishment of a joint line and continuum Monte Carlo code, however. The main issue for Monte Carlo radiative transfer is that the radiation fields always contain Monte Carlo noise, which leads to slow convergence. In particular, due to the random nature of the propagation direction of the photon packages, some cells are visited less frequently than others, which results in a slower convergence rate. A second challenge is that negative opacities can be encountered in non-LTE line radiative transfer. Indeed, inverted level populations can lead to net negative opacity coefficients and stimulated emission, and standard Monte Carlo radiative transfer methods cannot handle this. 

{\sc SKIRT}\footnote{The open-source {\sc SKIRT} code is registered in the ASCL with the code entry ascl:1109.003 and is hosted at \url{www.github.com/SKIRT/SKIRT9}. Documentation and other information can be found at \url{www.skirt.ugent.be.}} \citep{Baes2011, Camps&Baes2015, Camps2020SKIRT9} is a state-of-the-art 3D Monte Carlo code. Originally set up as a dust radiative transfer code, it has now evolved to a more generic radiative transfer code that also includes capabilities beyond dust radiative transfer \citep[e.g.,][]{Camps2021, 2023MNRAS.521.5645G, Bert2023} and is widely applied to generate synthetic ultraviolet (UV)--mm broadband fluxes, images, SEDs, and polarization maps for simulated galaxies and tori of active galactic nuclei \citep[AGNs, e.g.,][]{2012MNRAS.420.2756S, Stalevski2023, 2016MNRAS.462.1057C, 2022MNRAS.512.2728C, 2017MNRAS.470..771T, Granato2021, Kapoor2021, 2021A&A...653A..34V, Lahen2022, Ana2022, 2023MNRAS.519.2475H}. However, non-LTE line radiative transfer calculations have not been implemented in {\sc SKIRT} so far.

Thanks to several characteristics of the code already implemented, the extension of {\sc SKIRT} to a joint non-LTE line and dust continuum radiative transfer code is possible. To deal with Monte Carlo noise and poor convergence of the radiation field, {\sc SKIRT} is equipped with a hybrid parallelization strategy which allows for the efficient launching of a huge number of photon packets \citep{Verstocken2017, Camps2020SKIRT9}. Various Monte Carlo acceleration and variance-reduction techniques have been implemented to increase the number of photon packets crossing each cell and to stimulate the absorption rates in less frequently visited cells \citep[e.g.,][]{Baes2011, Baes2016, 2013ARA&A..51...63S, Camps2020SKIRT9}. To enable negative opacities in the Monte Carlo loop, the technique of explicit absorption has been introduced \citep{Baes2022}. This technique treats absorption along any photon packet path deterministically, while only scattering is treated stochastically. It boosts the efficiency of the Monte Carlo radiative transfer loop and is fully compatible with the negative opacities corresponding to stimulated emission.  

In this paper, we present the implementation of non-LTE line radiative transfer in {\sc SKIRT} and thus its extension to a self-consistent dust and non-LTE line radiative transfer code. This paper is organized as follows. Section~{\ref{Method}} describes the procedure of the self-consistent non-LTE line and dust radiative transfer calculation. In Section~{\ref{Benchmarks}}, we investigate whether the Monte-Carlo-based non-LTE line radiative transfer implemented in {\sc SKIRT} can achieve accuracy comparable to other codes using the standard benchmarks of \citet{VanZadelhoff_2002A&A}. 
Section~{\ref{Application}} demonstrates mock observations for CO lines against the dust continuum based on a 3D hydrodynamic simulation of a multi-phase AGN torus and investigates the effect of dust attenuation on CO lines. Section~{\ref{Conclusion}} discusses the characteristics, strengths, and limitations of our self-consistent dust and non-LTE line radiative transfer code and presents our conclusions.


\section{Dust and non-LTE line radiative transfer in SKIRT}
\label{Method}

{\sc SKIRT} is currently primarily a dust radiative transfer code, but is ideally suited to be extended for self-consistent dust and non-LTE RT. Indeed, some important features have already been introduced, such as Doppler shifts of both the emitting sources and the medium, and iterative schemes for the calculation of the radiation field and the dust temperature distribution \citep{Camps2020SKIRT9}.
The modified Monte Carlo radiative transfer scheme of explicit absorption has also been implemented in the latest version of {\sc SKIRT} \citep{Baes2022}. This scheme allows us to handle negative opacities by separately operating the absorption and stimulated emission numerically and the scattering process stochastically. Thanks to these features already implemented, the extension to self-consistent dust and non-LTE line radiative transfer only requires the implementation of the emissivity, the opacity, and the line profile of atoms and molecules, and the framework to calculate the level populations (see Appendix~{\ref{Ap: General formalization}} for a formal summary).

\begin{figure}
\includegraphics[width=0.45\textwidth]{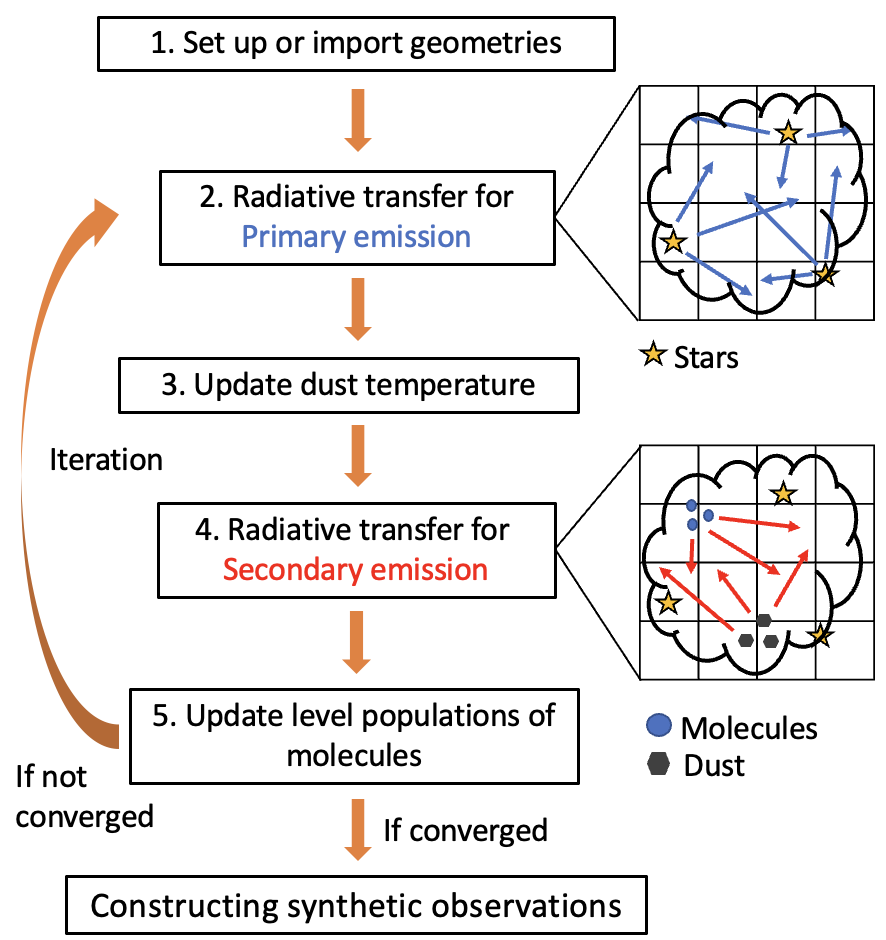}
\caption{Execution flow of self-consistent dust and non-LTE line radiative transfer calculation in {\sc SKIRT} and the schematic pictures of the radiative transfer of primary and secondary emission. Blue and red arrows represent primary and secondary emission, respectively.}
\label{fig:ExecutionFlow}
\end{figure}

We have set up an algorithm for iteratively performing the two-step radiative transfer calculations, updating dust temperatures, and solving statistical equilibrium equations.
Figure \ref{fig:ExecutionFlow} illustrates the execution flow of the dust and non-LTE radiative transfer in {\sc SKIRT}.
The first step consists of setting up the input parameters for the model. In most cases, this comes down to importing physical values from a hydrodynamic simulation snapshot. The physical properties that the radiative transfer routine requires are the number density of molecules or atoms $n_\mathrm{mol}$, the number densities of collision partners $n_{\text{p}}$, the kinetic gas temperature $ T_{\text{kin}}$, the microturbulence $v_{\text{turb}}$, and the densities and optical properties of the different dust species. It is also possible to use any analytical or semi-analytical geometries built into {\sc SKIRT}, including shells, exponential discs, smooth and clumpy tori, etc. \citep{Baes&Camps2015}. We also need to choose an appropriate grid to discretize the medium \citep{Camps2013, Saftly2013, Saftly2014} and to set the different wavelength grids to be used in the different phases of the simulation \citep[see][for details]{Camps2020SKIRT9}. For the line RT, we need to include a finely discretized grid around the wavelength of each transition line.

In the second step, we launch photon packets in the so-called primary emission cycle. {\sc SKIRT} works with two types of photon packets: primary and secondary emission photon packets. Primary emission photon packets are launched from primary sources, which are sources for which the emission spectrum and luminosity are fixed at the start of the simulation. This includes stars or stellar populations, 
AGNs, the CMB, or a user-defined background source. Secondary emission photon packets are those launched from sources for which the emission profile depends on the local radiation field, such as the dust and gas medium. For estimating the radiation fields corresponding to primary emission, a large number of primary emission photon packets are launched, with a wavelength, initial position, and propagation direction determined by the source characteristics.
The photon packets interact with the dust grains and the atoms and molecules in the medium. The optical properties of the dust grains are independent of the radiation field, but those of the atoms and molecules are not, as these depend on the radiation-field-dependent level populations. The initial level populations of atoms and molecules are assumed in LTE, and the initial opacity and emissivity are calculated.

When the primary emission cycle has terminated, we can, in every cell of our grid, calculate the radiation field from primary emission. we accomplish this calculation by averaging the intensity of all photon packets passing through each cell, as opposed to just considering information at the cell centers. Subsequently, we determine the temperature distribution of the different dust grains. The user can choose whether to use local thermal equilibrium or stochastic heating for the dust temperature distribution and the corresponding dust emission spectra \citep{Camps2015Dust}.

The fourth step consists of the radiative transfer corresponding to secondary sources, which in the present case comes down to photon packages from the dust and gas. 
Similar to the second step, a large number of photon packets are launched, and each of them is followed through the medium. 
Contrary to the primary emission cycle, the assigned energy carried by each secondary photon packet is variable in every iteration of the simulation, since it depends on the emission spectrum of the medium, which in turn depends on the level populations of atoms and molecules and the dust temperature distribution. For a given medium cell, the total secondary luminosity is the sum of the total dust luminosity and the joint luminosity corresponding to the spontaneous emission of all line transitions. For a given transition line, the total luminosity is given by
\begin{equation}
  L_{\text{line}} 
  = \frac{hc}{\lambda_{\text{line}}}\, A_{ul}\, n_u\,  V_{\text{cell}},
  \label{eq: Luminosity}
\end{equation}
where $\lambda_{\text{line}}$, $A_{ul}$, $n_u$, and $V_{\text{cell}}$ are the rest wavelength of the transition line, the Einstein coefficients for spontaneous emission from the upper level $u$ to lower level $l$, the number density of the upper level of the atoms or molecules, and the volume of the cell, respectively. $h$ and $c$ are the Planck constant and the speed of light, respectively. The wavelength of each photon packet $\lambda$ is determined according to the Gaussian distribution in the rest frame of the cell and adjusted for any Doppler shifts caused by the bulk motion at the cell. The probability function of the Gaussian distribution, which is the same as the line profile $\phi_{\text{line}}$, is expressed as 
\begin{equation}
  P_{\text{line}}(\lambda) = \phi_{\text{line}}(\lambda) = \frac{1}{\sqrt{2\pi}\, \sigma_{\text{line}}} \exp\left(-\frac{(\lambda-\lambda_{\text{line}})^2}{2\, \sigma_{\text{line}}^2}\right). \label{eq:Gaussian_profile}
\end{equation}
The wavelength dispersion $\sigma_{\text{line}}$ is given by
\begin{equation}
  \sigma_{\text{line}} = \frac{\lambda_{\text{line}}}{c}\, \sqrt{\frac{ k T_{\text{kin}}}{m_{\text{s}}} + \frac{v_{\text{turb}}^2}{2}}, \label{eq:velocity_dispersion}
\end{equation}
where $k$ and $m_\mathrm{s}$ are the Boltzmann constant and the atomic or molecular mass, respectively. The first term and the second term under the square root represent the sound speed of the thermal motion of the gas and the assumed non-thermal motion of clouds in each grid, respectively.
After launching all secondary source photon packets and following their life cycle through the medium, we can calculate the radiation field corresponding to the secondary emission.

The fifth step corresponds to the most important extension of the {\sc SKIRT} code. Once the radiation fields from primary $J_\lambda^\mathrm{prim}$ and secondary emission $J_\lambda^\mathrm{second}$ in each cell are estimated, we calculate the combined radiation field,
\begin{equation}
J_\lambda = J_\lambda^{\text{prim}} + J_\lambda^{\text{second}},
\end{equation}
and we determine the mean intensity at each line transition,
\begin{equation}
    \bar{J}_{\text{line}} = \int J_\lambda \,\phi_{\text{line}}(\lambda) \, {\text{d}}\lambda.
     \label{eq: mean intensities for transitions}
\end{equation}
To obtain the level populations, we assume that the timescale to the equilibrium of transitions in the level populations is shorter than the dynamical timescale of the system. We can formulate the statistical equilibrium equations as follows,
\begin{multline}
 \sum^{N_\mathrm{lp}}_{j<i}  n_iA_{ij} - \sum^{N_\mathrm{lp}}_{j>i}  n_jA_{ji}
 + \sum^{N_\mathrm{lp}}_{j\neq i} \Big[ (n_iB_{ij} - n_jB_{ji}) \bar{J}_{\mathrm{line}}\Big]
 \\
 + \sum^{N_\mathrm{lp}}_{j\neq i}\sum^{N_\mathrm{p}}_{p}\Big[n_iC_{ij}(n_\mathrm{p},T_\mathrm{kin}) - n_jC_{ji}(n_\mathrm{p},T_\mathrm{kin})\Big]=0,
 \label{eq:Statistical_equation}
 \end{multline}
where $N_\mathrm{lp}$ and $N_\mathrm{p}$ are the numbers of energy states and collision partners, respectively. $B_{ul}$, $B_{lu}$, and $C_{ij}$ are the Einstein coefficients for stimulated emission and absorption, and the collisional coefficients, respectively (see also Appendix \ref{Ap: Collisional coefficients}).
The upper and lower terms on the left side represent radiative and collisional transition rates, respectively. Equation~(\ref{eq:Statistical_equation}) forms a matrix equation with the level populations $n_i$ as the unknowns. To solve this matrix equation, we use the LU decomposition method. Once the level populations are updated, the absorption coefficients and the luminosity of the atoms and molecules are also updated (see Appendix~{\ref{Ap: General formalization} and Equation~(\ref{eq: Luminosity}), respectively).

After finishing the fifth step, we return to the second one, the primary emission radiative transfer calculations. Indeed, with the level populations of atoms and molecules updated, the absorption coefficients also changed. This naturally leads to an iterative cycle of calculations from step 2 to step 5 until the solution is converged. To accelerate the iterative calculations, Ng acceleration \citep{Ng1974} is implemented in {\sc SKIRT} but not used in the default configuration.

We have implemented two convergence criteria in {\sc SKIRT}, and depending on their specific use case, users can select which one they use (or they can implement a different one). The first one employs the local convergence parameter $\epsilon_m$, given by
 \begin{equation}
  \epsilon_m = \frac{1}{N_\mathrm{lp}}\, \sum_i \ \left| \, \frac{n_{i,m}^{(k)} - n_{i,m}^{(k-1)}}{n_{i,m}^{(k-1)}}\,\right| \label{eq:local convergence},
\end{equation}
where $n_{i,m}^{(k)}$ represents the number density at the energy states $i$ in cell number $m$ at iteration stage $k$. 
Our default convergence criterion is that $\epsilon_m$ must be below 0.05 for 99.9\% of all cells in the simulation volume, but the fraction of converged cells and the threshold value can be defined by users.

An alternative convergence criterion implemented employs the total convergence parameter $\epsilon_{\text{total}}$, given by

\begin{equation}
  \epsilon_{\text{total}} 
  = 
  \max_i\ \left|\,\frac{n_i^{(k)} - n_i^{(k-1)}}{n_i^{(k-1)}}\, \right| 
  \label{eq:total convergence},
\end{equation}
where $n_i^{(k)}$ represents the total number density in the entire simulation volume at the energy state $i$ at iteration stage $k$. Our default setting is that iterations have converged when  $\epsilon_{\text{total}}$ is below $10^{-4}$, but, again, it can be chosen by the user. We note that the total convergence parameter minimizes the contribution of Monte Carlo noise and is, therefore, suitable as a convergence criterion in Monte Carlo radiative transfer simulations.

Finally, when convergence has been reached, we proceed to the final step in our Monte Carlo radiative transfer algorithm, i.e., the construction of synthetic observations.
We again perform radiative transfer calculations for primary and secondary emission in steps 2 and 4 based on the converged absorption coefficients and emissivity of dust, atoms, and molecules.
Photon packets escaping from the system are detected by synthetic instruments located at arbitrary viewing points with user-defined characteristics such as field-of-view and spatial and wavelength resolution. For the specific case of joint dust and non-LTE line RT, we typically use instruments with a wide wavelength range consisting of relatively broad wavelength bins, but with narrow wavelength bins around the line centers. In order to boost the signal-to-noise in these narrow bins around the line centers, composite biasing in wavelength space can be employed \citep{Baes2016, Camps2020SKIRT9}.

At this moment we have implemented non-LTE radiative transfer for two atoms and four molecules (C, C$^+$, CO, OH, H$_2$, and HCO$^+$) in {\sc SKIRT}.
Relevant atomic and molecular properties are taken from the Leiden Atomic and Molecular Database\footnote{\url{https://home.strw.leidenuniv.nl/~moldata}} and the BullDog Database\footnote{\url{https://www.physast.uga.edu/amdbs/excitation/}}.


\section{Benchmarks}
\label{Benchmarks}

To validate the non-LTE line radiative transfer in {\sc SKIRT}, we run the code on four well-defined test problems presented by \citet{VanZadelhoff_2002A&A}. We compare our results with their reference results and those obtained by a modern, pure ray-tracing non-LTE line radiative transfer code, {\sc MAGRITTE} \citep{Frederik2020}. In the {\sc MAGRITTE} simulations, the convergence criterion is that $\epsilon_m<10^{-10}$ for 99.5\% of all cells in the simulation volume through all problems.
The parameter files of these benchmarks for {\sc SKIRT}\footnote{\url{https://skirt.ugent.be/root/_zadelhoff2002.html}} are publicly available.


\subsection{Problem 1a/1b}
\label{Problem 1a/1b: a simple 2-level molecule}

The first set of two models is originally taken from \citet{Dullemond2000} and corresponds to a simple spherically symmetric shell without radial velocities. The shell is populated by molecular hydrogen and by an artificial two-level molecule specified by
\begin{gather}
\Delta E = 6.0 \ \ \mathrm{cm^{-1}}, \\
g_0,\, g_1 = 1, \, 3,   \\
A_{10} = 1.0 \times 10^{-4} \ \ \mathrm{s^{-1}}, \\
K_{10} = 2.0 \times 10^{-10} \ \ \mathrm{cm^3 \ s^{-1}},\\
M_\mathrm{mol} = 14.67,
\end{gather}
where $g_0$, $g_1$, and $K_{10}$ are the degeneracy of the lower and upper energy levels and the collisional de-excitation coefficient, respectively. $\Delta E$ and $M_\mathrm{mol}$ are the energy gap between the two energy levels of the molecule and mass expressed in units of the hydrogen atom mass, respectively.
The molecular mass is determined so that the thermal velocity is 150 $\mathrm{m\,s^{-1}}$ at the kinetic temperature of 20 K. 

The artificial molecules and the molecular hydrogen, which acts as the collisional partner, are distributed in a spherical shell with inner radius $r_0 = 1.0 \times 10^{15}$~cm and outer radius $r_1 = 7.8 \times 10^{18}$~cm. The radial profile of the molecular hydrogen number density is given by a power law of
\begin{equation}
    n_\mathrm{H_2} (r)= n_\mathrm{H_2} (r_0)
    \left(\frac{r}{r_\mathrm{0}} \right)^{-2} ,
\end{equation}
where $r$ is the radius, and $n_\mathrm{H_2} (r_0)=2.0 \, \times \, 10^{7} \, \mathrm{cm^{-3}}$. The kinetic temperature and micro-turbulence of the molecules are constant at $20$ K and zero, respectively. The abundances of the artificial molecules in problems 1a and 1b are $X_{\text{a}} = 10^{-8}$ and $X_{\text{b}}=10^{-6}$, respectively, and the total optical depths at the line center are $60$ and $4800$, respectively. The shell is illuminated by the external radiation from the CMB.

For the {\sc SKIRT} calculations, we have set up a spherical grid with 50 and 500 logarithmically spaced shells for problems 1a and 1b, respectively (the dependency of the results of problem 1b on the number of shells is investigated in Section \ref{subsec: Requirement of cell numbers in SKIRT}). We adopted $10^8$ photon packets for these problems, and the convergence criterion is that $\epsilon_m<0.001$ for 99 \% of all cells in the simulation volume.
To reach convergence in the level populations, {\sc SKIRT} requires 42 and 694 iterations in problems 1a and 1b, respectively.
On the other hand, {\sc MAGRITTE} needs 34 and 379 iterations in problems 1a and 1b, respectively.

\begin{figure}[t]
\includegraphics[width=0.5\textwidth]{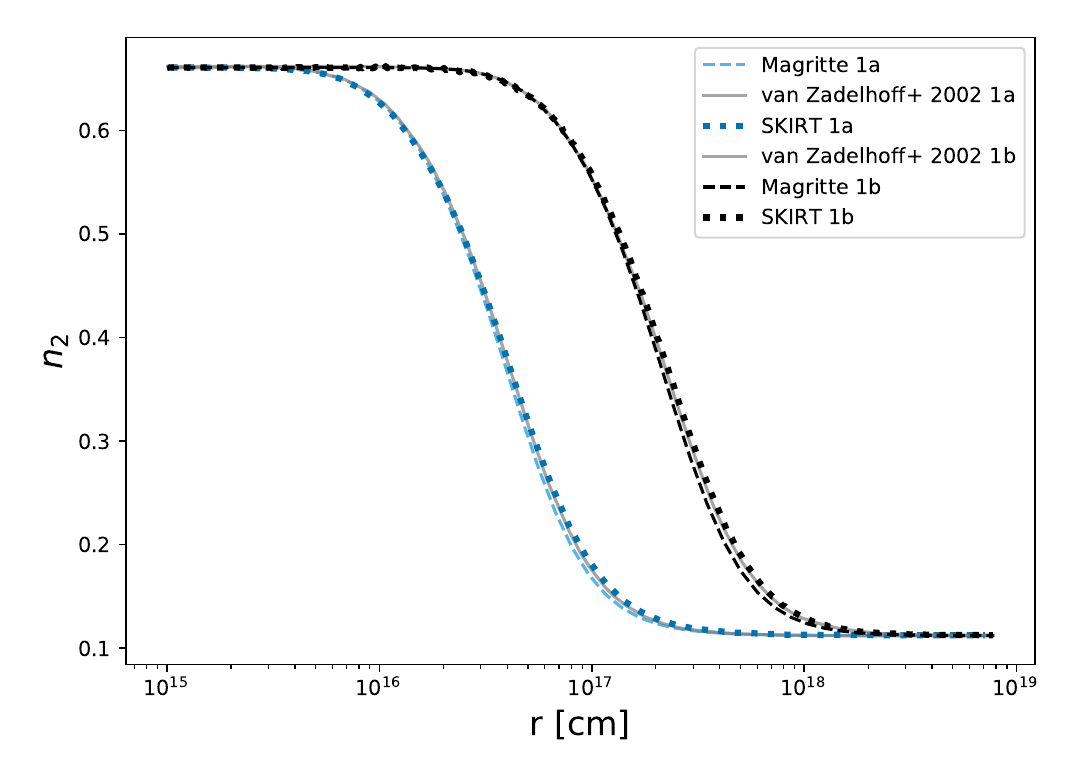}
\caption{Comparison of the relative population at the upper level obtained with {\sc SKIRT} (dots) and {\sc MAGRITTE} (dashed lines) and the reference results of \citet{VanZadelhoff_2002A&A} (grey solid lines) for benchmark problem 1a (blue lines) and 1b (green lines). The level population is normalized by the total number density. The differences of {\sc MAGRITTE} and {\sc SKIRT} results relative to the reference results are shown in the bottom panel.}

\label{Fig: Benchmark Problem 1a/b}
\end{figure}

Figure \ref{Fig: Benchmark Problem 1a/b} shows the comparison of the level populations among the different codes used in the \citet{VanZadelhoff_2002A&A} reference study (solid lines), {\sc SKIRT} (dots) and {\sc MAGRITTE} (dashed lines) for problems 1a and 1b. In these problems, the level populations at small radii are collisionally determined, while at large radii they are radiatively determined by the CMB.
Since the shell in problem 1b is more optically thick than the one in problem 1a, the CMB radiation cannot penetrate deeply into the gas shell and as a result, the level populations of problem 1b are collisionally determined out to the larger radii.
The level populations obtained with {\sc SKIRT} agree with the reference results within the relative difference of 2 and 5 \% in problems 1a and 1b, respectively, while those obtained with {\sc MAGRITTE} agree within the relative difference of 2 \% in both problems.


\subsection{Problem 2a/2b: a collapsing cloud in HCO$^+$}
\label{2a/2b}

The second set of two models is based on an inside-out collapse model presented by \citet{Shu1977}. These problems again assume a 1D spherical shell geometry, now with inner radius $r_0 = 1.0 \times 10^{16}$~cm and outer radius $r_1 = 4.6 \times 10^{17}$~cm, and with density, velocity, and temperature gradients and a non-constant microturbulence.
The radial profiles of those physical values are shown in Fig.~2 of \citet{VanZadelhoff_2002A&A}.
The shell contains HCO$^+$ molecules and H$_2$ molecules which act as collisional partners. The $\mathrm{HCO^+}$ abundances in problems 2a and 2b are $X_{\text{a}} = 10^{-8}$ and $X_{\text{b}} = 10^{-6}$, respectively.

For the {\sc SKIRT} calculations of these two benchmark problems, we have set up a 3D cubic box with $100^3$ cuboidal cells. The distribution of cells along each axis is chosen according to a power-law distribution to ensure smaller cells near the inner regions of the shell. 
For problems 2a and 2b, we use $2.2 \times 10^8$ and $10^9$ photon packets, respectively. We set the convergence criterion as that $\epsilon_m<0.02$ for 99.95 \% of all cells.
To reach the convergence criterion, {\sc SKIRT} requires 22 and 94 iterations in problems 1a and 1b, respectively.
On the other hand, {\sc MAGRITTE} needs 33 and 110 iterations in problems 1a and 1b, respectively.

Figure \ref{Fig: Benchmark Problem 2a/b} shows the comparison of the $J=1$ and $J=4$ level populations of $\mathrm{HCO^+}$ among the different codes used in the \citet{VanZadelhoff_2002A&A} reference study  (grey dashed lines),  {\sc SKIRT} (blue solid lines with dots) and {\sc MAGRITTE} (yellow solid lines) for problems 2a and 2b.
In both problems, the {\sc SKIRT} results agree with those obtained with {\sc MAGRITTE} within 20\%. Given that the level populations obtained using different codes in \citet{VanZadelhoff_2002A&A} show relative differences of up to 20\%, we argue that these tests validate {\sc SKIRT} as a non-LTE line radiative transfer code.

\begin{figure}
\includegraphics[width=0.5\textwidth]{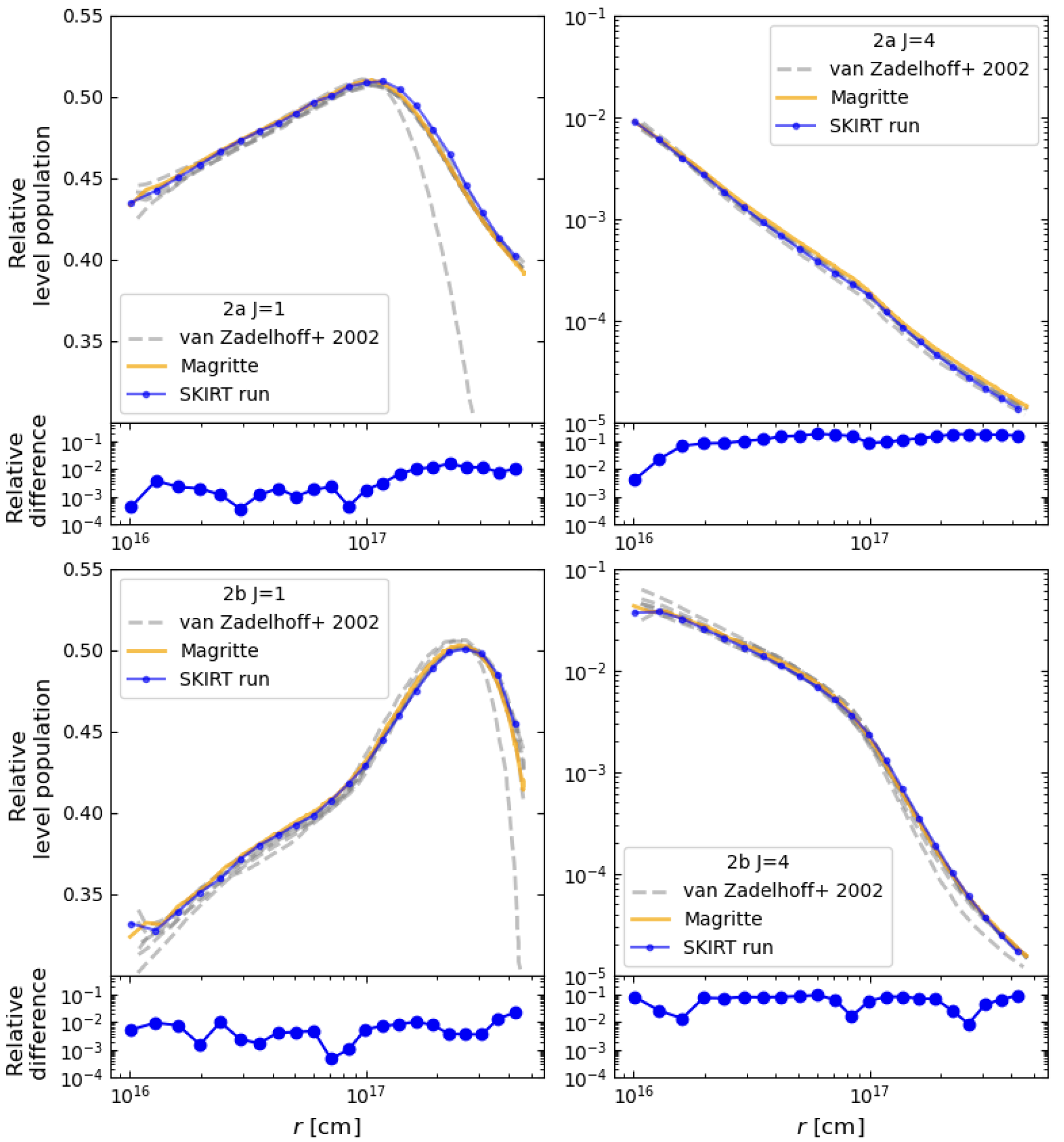}
\caption{Comparison of the relative $\mathrm{HCO^+}$ level population for benchmark problem 2a (upper panels) and 2b (bottom panels) at $J=1$ (left) and $J=4$ (right) obtained with {\sc SKIRT} (blue solid lines with dots) and {\sc MAGRITTE} (yellow solid lines) and the reference results of \citet{VanZadelhoff_2002A&A} (grey dashed lines). The level population is normalized by the total number density.
The relative differences between the {\sc MAGRITTE} and {\sc SKIRT} results are shown in the lower subpanels.}
\label{Fig: Benchmark Problem 2a/b}
\end{figure}

\subsection{Convergence of the radiation field in {\sc SKIRT}: tests based on problem 1b}
\label{sec: Limitations of SKIRT}

The iterative calculation in non-LTE line radiative transfer calculations plays an essential role in sharing radiation throughout the entire simulation box.
However, if the optical thickness per cell becomes too large, photon packets are primarily absorbed within a single cell, which results in radiation fields that are no longer shared between neighboring cells. The solution consists of increasing the number of cells and/or the number of photon packets in the simulation, but this immediately affects the run-time and memory requirements.

In this subsection, we investigate two critical parameters, the number of grid cells and the number of photon packets, that determine the stability of the radiation field.
We use problem 1b of \citet{VanZadelhoff_2002A&A} for these tests; this problem is ideally suited for this test because of the combination of simplicity (we can use a 1D spherical grid in which all cells are concentric shells) and high optical thickness ($\tau \sim 4800$ at the line center).


\subsubsection{The number of cells}
\label{subsec: Requirement of cell numbers in SKIRT}

The limitation of the optical thickness in Monte Carlo dust radiative transfer calculations with {\sc SKIRT} is $\tau<75$ \citep{Camps2018} while $\tau<100$ is required for the ray-tracing method \citep{Hogerheijde&vanderTak2000}. However, it is not clear whether non-LTE Monte Carlo radiative transfer calculations also require $\tau<75$ for sharing the radiation fields.
Here, we investigate the limitations of the optical thickness per cell by performing {\sc SKIRT} on problem 1b with various numbers of grid cells.

\begin{figure}
\includegraphics[width=0.5\textwidth]{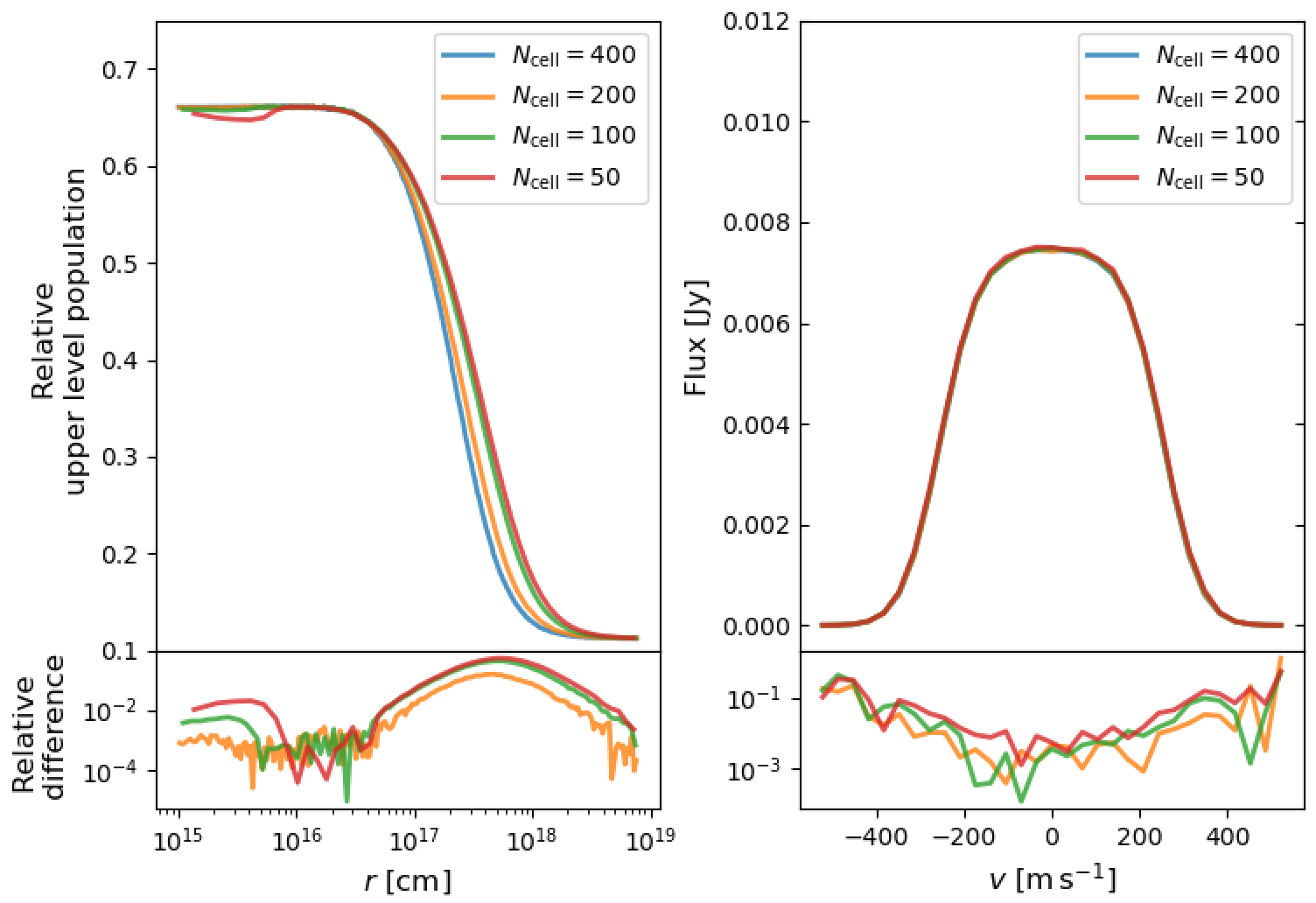}
\caption{Convergence tests for benchmark problem 1b for calculations with different cell numbers, as indicated in the top right corner of the left panel. \textit{Left}: relative populations at the upper level. \textit{Right}: observed line profiles corresponding to a distance of 1 Mpc from the sphere's center.
In both cases, the relative differences between each simulation and the reference simulation with 400 cells are shown in the lower panels.}
\label{Fig: Requirement of cell numbers}
\end{figure}

The left panel of Figure \ref{Fig: Requirement of cell numbers} compares the relative populations at the upper level of the artificial two-level molecule corresponding to calculations with different numbers of grid cells ($N_{\text{cell}} = 50$, 100, 200, and 400), and we consider the simulation with the highest number of cells as our ground truth result. We use 750 iterations, and the same number of photon packets in each iteration, $10^7$. 
The populations corresponding to the calculations with smaller numbers of cells show larger deviations from the ground truth results, which implies that if the simulation box is not sufficiently divided, the radiation fields are no longer shared. For the calculation with $N_\mathrm{cell}=400$, the largest optical depth in any cell is $73.7$, which suggests that non-LTE radiative transfer calculations in {\sc SKIRT} require a similar optical-depth-per-cell constraint as dust radiative transfer calculations to reproduce accurate level populations.

On the other hand, the right panel of Fig. \ref{Fig: Requirement of cell numbers} shows the observed line profiles of the two-level molecule of problem 1b at the 1.6666 mm transition, for different {\sc SKIRT} simulations with $N_\mathrm{cell}=50$,\ 100,\ 200,\ and 400.
The line profiles are not significantly different since only photon packets from the surface layer of the sphere can reach the observer because of the large optical thickness.


\subsubsection{The number of photon packets}
\label{subsec: Requirement of photon numbers}

To investigate the required number of photon packets to solve problem 1b, we run a suite of {\sc SKIRT} simulations with various numbers of photon packets. We use 500 cells and 750 iterations in each simulation.

The left panel of Fig. \ref{Fig: Requirement of photon numbers} compares the relative population at the upper level of the two-level molecule from problem 1b for {\sc SKIRT} simulations with various numbers of photon packets ($N_{\text{p}} = 1\times10^4$, $4\times10^5$, and $1\times10^7$). The solution corresponding to $N_{\text{p}} = 1\times10^7$ is considered to be our ground truth.  While the level population of the calculations with $N_\mathrm{p}=4\times10^5$ is consistent with those with $N_\mathrm{p}=1\times10^7$, that with $N_\mathrm{p}=1\times10^4$ is not. The reason for this mismatch is that, for a small number of photon packets per cell (here $N_\mathrm{p}/500$), we cannot accurately express the shape of the line profile and the spectrum of the radiation field in each cell.

The right panel of Fig. \ref{Fig: Requirement of photon numbers} shows the comparison of the observed line profiles of problem 1b for the same set of three {\sc SKIRT} simulations. Noise is noticeable in the line profiles corresponding to the calculation with $N_\mathrm{p}=4\times10^5$. To calculate accurate radiation fields and produce correct level populations, we suggest at least the total number of photon packets of $N_\mathrm{p}=800 \times N_\mathrm{cell} \times N_\mathrm{line}$, where $N_\mathrm{line}$ is the number of the transition lines of molecules or atoms in the calculations.


\begin{figure}
\includegraphics[width=0.5\textwidth]{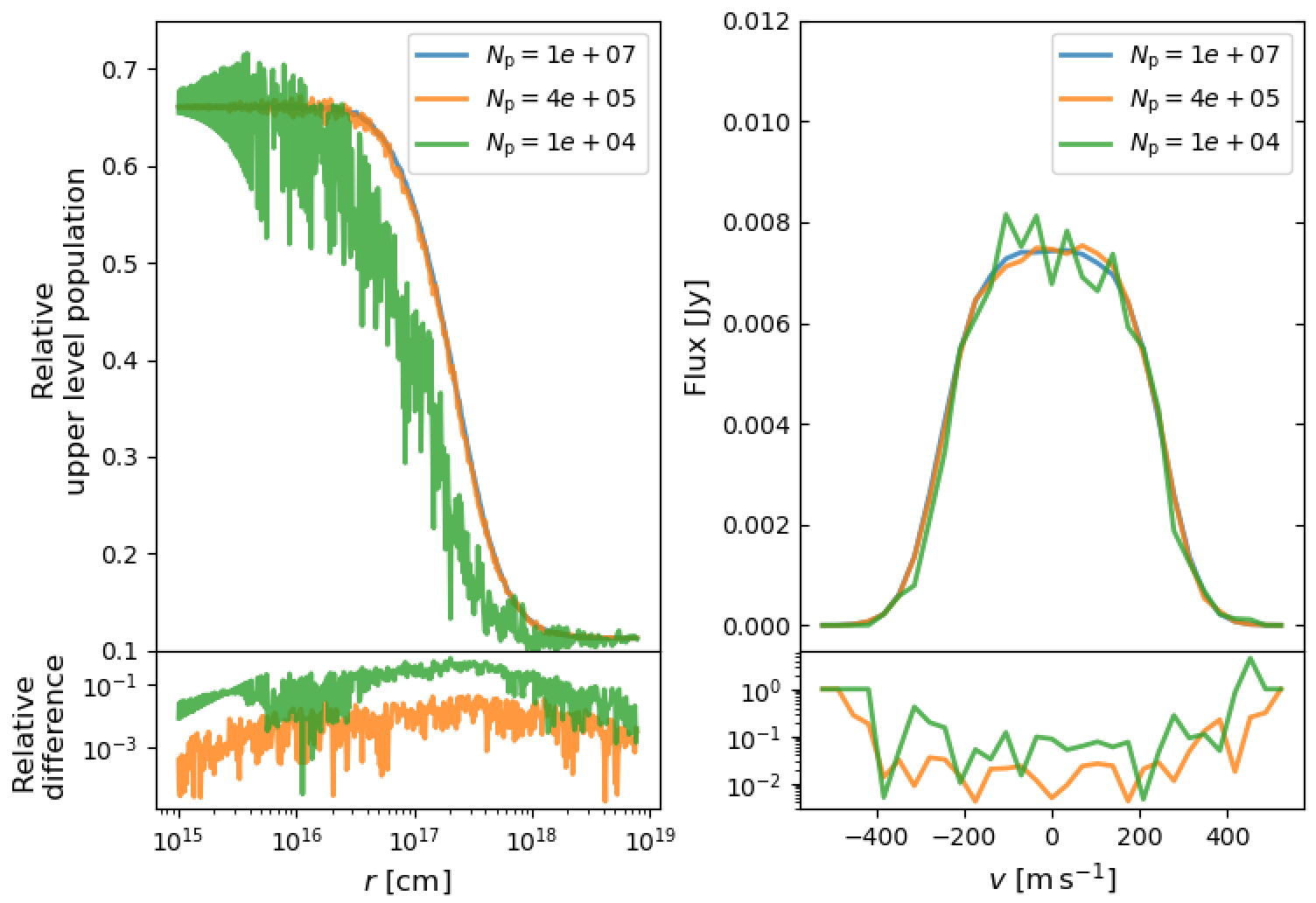}
\caption{Convergence tests for benchmark problem 1b for calculations with different photon packet numbers, as indicated in the top right corner of the left panel. \textit{Left}: relative populations at the upper level. \textit{Right}: observed line profiles corresponding to a distance of 1 Mpc from the sphere's center.
In both cases, the relative differences between each simulation and the reference simulation with $10^7$ photon packets are shown in the bottom panels.}
\label{Fig: Requirement of photon numbers}
\end{figure}


\section{Application to an AGN torus model}
\label{Application}

As a demonstration of the power of the new capabilities of {\sc{SKIRT}}, we perform self-consistent dust and non-LTE line radiative transfer calculations based on an AGN torus simulation.

\subsection{AGN torus model and SKIRT setup}
\label{Application: Wada2016 model}

\citet{Wada2016} modeled the structure of the AGN torus of the Circinus galaxy using a 3D Eulerian hydrodynamic simulation composed of a uniform Cartesian grid with $256^3$ cells. The simulation box is 32~pc on each side, corresponding to a cell resolution of 0.125~pc.
The model incorporates the physics coupling radiative processes and hydrodynamics: UV and X-ray radiation from the accretion disk drives outflows in the polar direction of the AGN and generates a geometrically thick torus with complex dynamical structures around the AGN. Furthermore, the radiation heats the ambient gas and affects the chemical structure, which results in the formation of multi-phase gas in the torus composed of molecular, atomic, and ionized phases. 
The abundances of 25 gas species including $\mathrm{H}$, $\mathrm{H_2}$, $\mathrm{H^+}$, $\mathrm{C}$, $\mathrm{C^+}$, and $\mathrm{CO}$ are calculated using the chemical network of \citet{Meijerink2005}. The dust-to-gas mass ratio is assumed to be $0.01$.
We assume that supernova explosions occur at random locations in the equatorial plane with a rate of 0.014 yr$^{-1}$, which helps to enhance the thickness of the torus \citep{Wada2009ApJ...702...63W}.  

The dust and CO emission of this AGN torus model has already been independently investigated by \citet{Wada2016, Wada2018a}, \citet{Uzuo2021}, and \citet{Matsumoto2022}. In these studies, the original data for the hydrodynamic simulation was reduced to half of the resolution due to the memory capacities of the radiative transfer codes used, and their simulations included uncertainties due to the reduced resolution. However, {\sc SKIRT} can keep the original spatial resolution due to the efficient memory distribution implemented in the code \citet{Camps2020SKIRT9}. In addition, {\sc SKIRT} can reconstruct the original data using hierarchical octree grids, recursively subdividing the simulation box into $2^4$ to $2^{8}$ cells in each axis. Consequently, the number of cells does not increase so much, though the original spatial resolution is retained. 

In our {\sc{SKIRT}} run we considered two sources of external radiation: the CMB and a central point source representing the AGN accretion disk. For this latter source, we assumed a power-law spectrum in the UV-to-optical wavelength range, which heats ambient dust.
We note that free-free and synchrotron radiation expected to be emitted near the accretion disk is not taken into account in our model, as the radiation is not regarded as the dominant source of the submillimeter continuum emission of AGN \citep[e.g.,][]{Tristram2022}.
We employ the \citet{Draine&Li2007} dust model with silicate, graphite, and polycyclic aromatic hydrocarbon (PAH) grains, and considered stochastic heating. The microturbulence in each cell was assumed to be constant at 5 $\mathrm{km\,s^{-1}}$.

To illustrate the computational cost of the {\sc{SKIRT}} simulations, we run two different simulations: (i)~a pure dust radiative transfer simulation with the accretion disk as the external source and (ii)~a self-consistent dust and CO line radiative transfer simulation with both the accretion disk and the CMB as external sources. We run the first simulation with $6\times10^{8}$ and $1.5\times10^{9}$ photon packets for the emission from the accretion disk and dust emission, respectively, and the second one with the additional $2.4\times10^{9}$ and $1.5\times10^{9}$ photon packets for CMB and the CO line emission, respectively. As a result, the second simulation adds 285 \% to the run times of each iteration in the first one. Moreover, the second simulation takes 10 times more iterations to reach convergence. Thus, the total run time of the second simulation increases as the iteration number and the number of total photon packets increase.
In Section~{\ref{Application: CO rotational lines and dust continuum}} we investigate the spectral features of the dust continuum and CO lines up to $J=20$.
 In Section~{\ref{Application: dust extinction effect on CO lines}} we investigate the dust extinction effects on the CO lines by comparing the results of the calculations with and without dust.

\begin{figure*}[h]
\includegraphics[width=\textwidth]{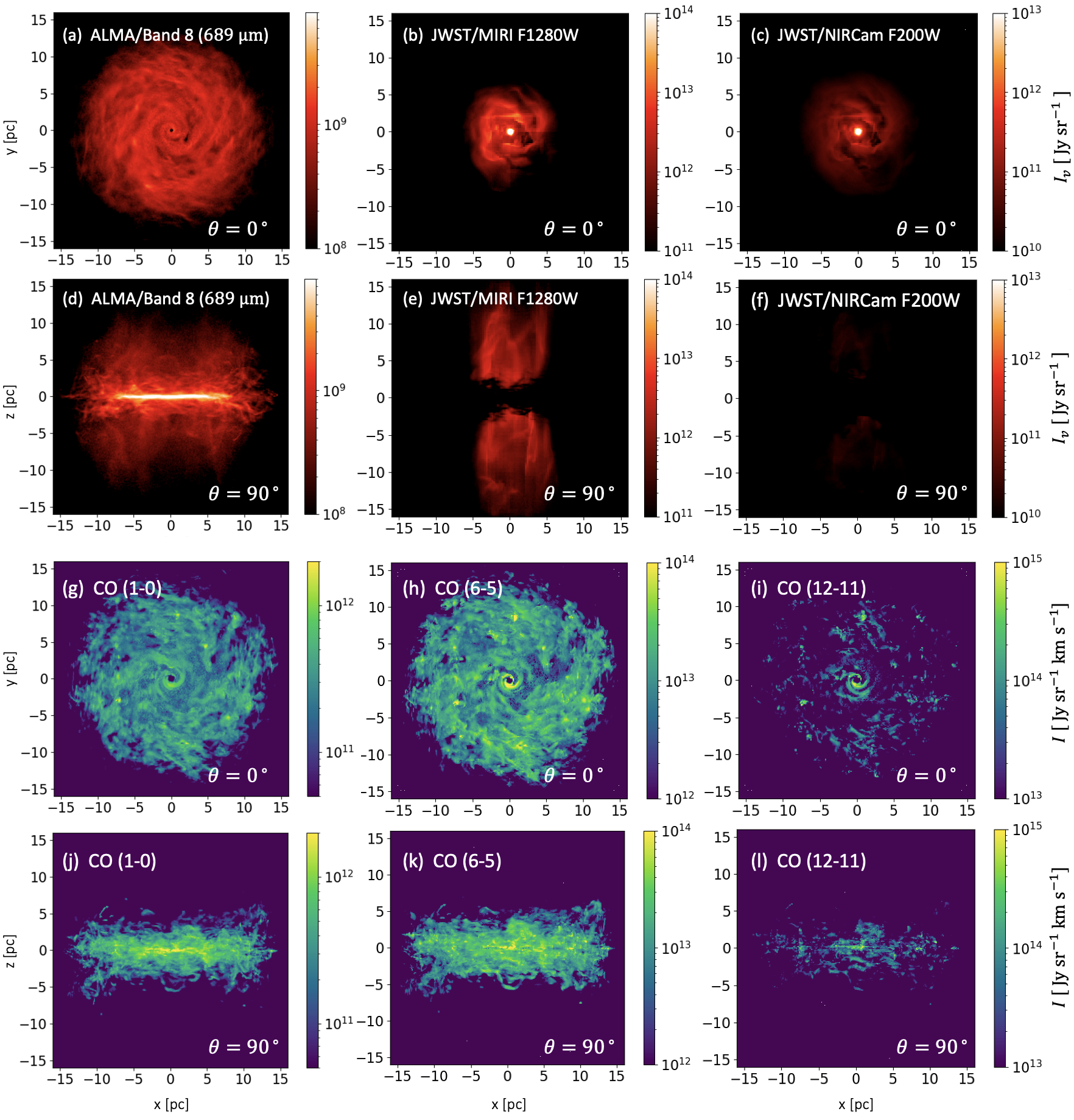}
\caption{Dust continuum images (first and second rows) and CO line emission maps (third and fourth rows) for the AGN torus model discussed in Section~{\ref{Application: Wada2016 model}}. Each image consists of $256\times256$ pixels with a pixel size of 0.125~pc. The images on the first and third rows are viewed face-on, and those on the second and fourth rows are viewed edge-on.}
\label{Fig: Circinus map}
\end{figure*} 

\subsection{CO rotational emission lines and dust continuum}\label{Application: CO rotational lines and dust continuum}

Figure \ref{Fig: Circinus map} shows the {\sc{SKIRT}} simulation results. The panels on the first two rows show dust continuum maps corresponding to three different bands: the far-infrared \textit{ALMA}/Band 8, the mid-infrared \textit{JWST}/MIRI F1280W band, and the near-infrared \textit{JWST}/NIRCam F200W band. The first and second rows correspond to face-on and edge-on views, respectively. 

While the far-infrared radiation is emitted only from the cold dust extending on the dense disk in the torus (panels~a and~d), the mid-infrared emission mainly comes from the inner region of the disk and the polar region of the torus (panels~b and~e).
Finally, the near-infrared emission can only be clearly detected in the innermost region of the AGN torus in the face-on orientation (panel~c). The emission is very faint in the edge-on view, as the emission is almost completely shielded by cold dust in the torus (panel~f). These results are qualitatively consistent with \citet{Schartmann2014}, in which the dust radiative transfer was performed with {\sc RADMC-3D} \citep{Dullemond2012} based on the hydrodynamic model presented by \citet{Wada2012}.

The panels on the third and fourth rows of Fig.~{\ref{Fig: Circinus map}} show integrated intensity maps of CO(1--0), CO(6--5), and CO(12--11) for face-on and edge-on views, respectively. Both the CO(1--0) and CO(6--5) emission extend to 10~pc on the vertical axis, tracing the geometrically thick molecular torus. Since the CO(6--5) emission originates from the molecular gas with kinetic temperatures of several tens K, it seems to trace a more puffed-up structure of the molecular torus compared to the CO(1--0) emission.
On the other hand, the CO(12--11) emission comes from gas with kinetic temperatures higher than 100 K in the inner-most region where the gas is heated by radiation from the accretion disk and the places where supernovae explode.

Figure~{\ref{Fig: Circinus SED Lines}} shows the total SED of the torus at various inclination angles ($\theta = 0^\circ, \ 45^\circ$, and $90^\circ$). The contribution of dust continuum emission, PAH band emission, the 9.7~silicate feature, and CO rotational lines are clearly visible. The dust emission in the near- to mid-infrared wavelength range is more and more attenuated as the inclination angle increases since this emission originates from the hot dust in the innermost region of the AGN torus and is shielded by cold dust in the outer region (see also Fig. \ref{Fig: Circinus map}).
The silicate feature is seen in emission at face-on and intermediate inclination, and in absorption in the edge-on view. At smaller inclination angles, higher-$J$ CO emission lines are stronger since we directly observe highly excited CO emission from the gas in the inner region of the AGN torus, and the high CO lines are attenuated by dust.

Figure~{\ref{Fig: Circinus Line profiles}} shows the integrated line profiles of CO(1--0), CO(6--5), CO(12--11), and CO(15--14) observed at various inclination angles ($\theta = 0^\circ, \ 45^\circ$, and $90^\circ$). As the inclination angle increases, the line profiles are broadened due to the Keplerian rotation of the AGN torus. In particular, the CO(1--0) emission line is observed as two separate profiles in the edge-on view, tracing the Keplerian motion of the cold gas. On the contrary, higher-$J$ lines are observed as asymmetrical profiles including broad components at the inclination angles of $\theta = 45^\circ$ and $90^\circ$, tracing the turbulent gas driven by radiation from the accretion disk and supernovae.

\begin{figure}
\includegraphics[width=0.5\textwidth]{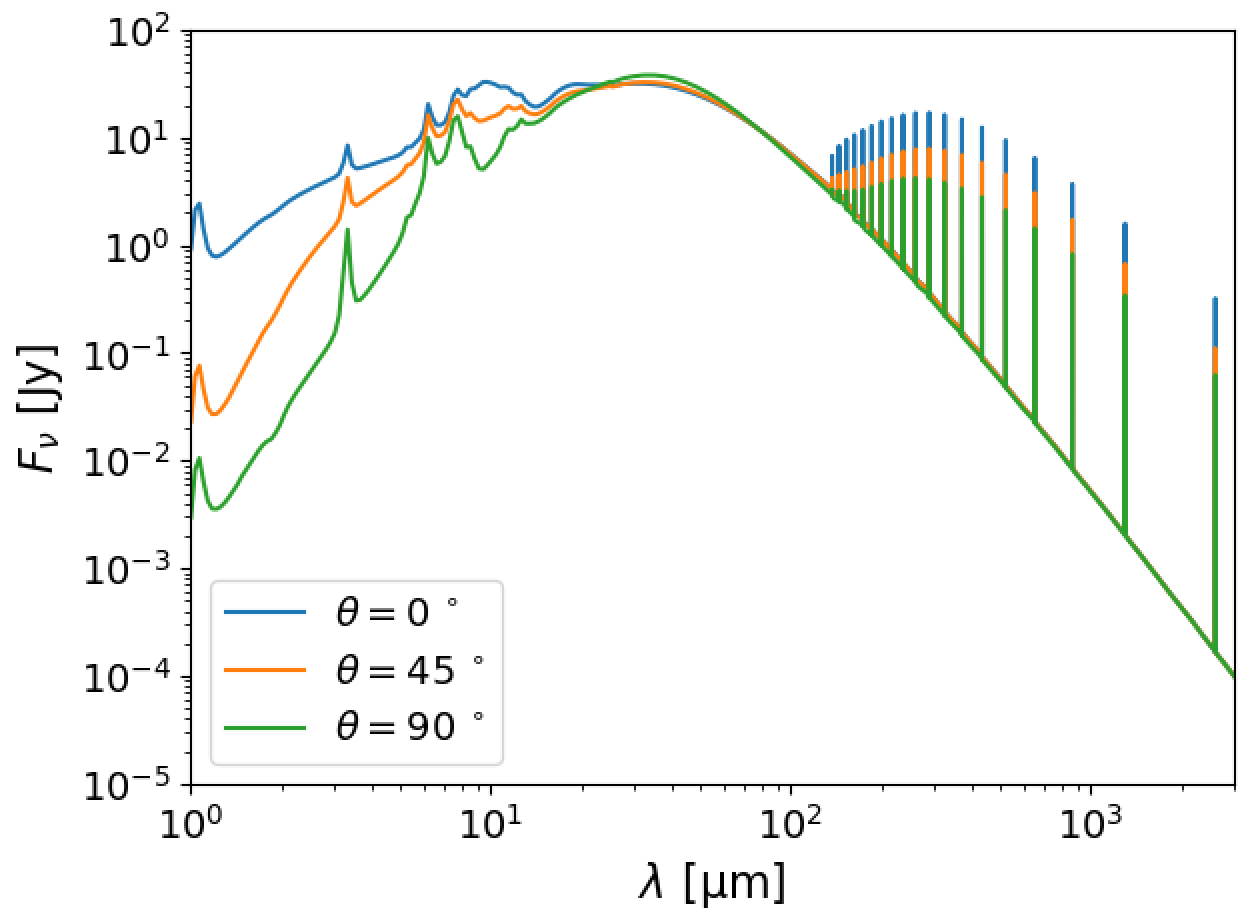}
\caption{Spectral energy distributions including CO rotational lines from $J=0$ to $J=19$ for the AGN torus simulation. The different lines correspond to SEDs seen at different inclination angles, as indicated in the legend. The observers are located at a fiducial distance of 4.2 Mpc, and a beam size of (32 pc)$^2$ is used.}
\label{Fig: Circinus SED Lines}
\end{figure}

\begin{figure}
\includegraphics[width=0.5\textwidth]{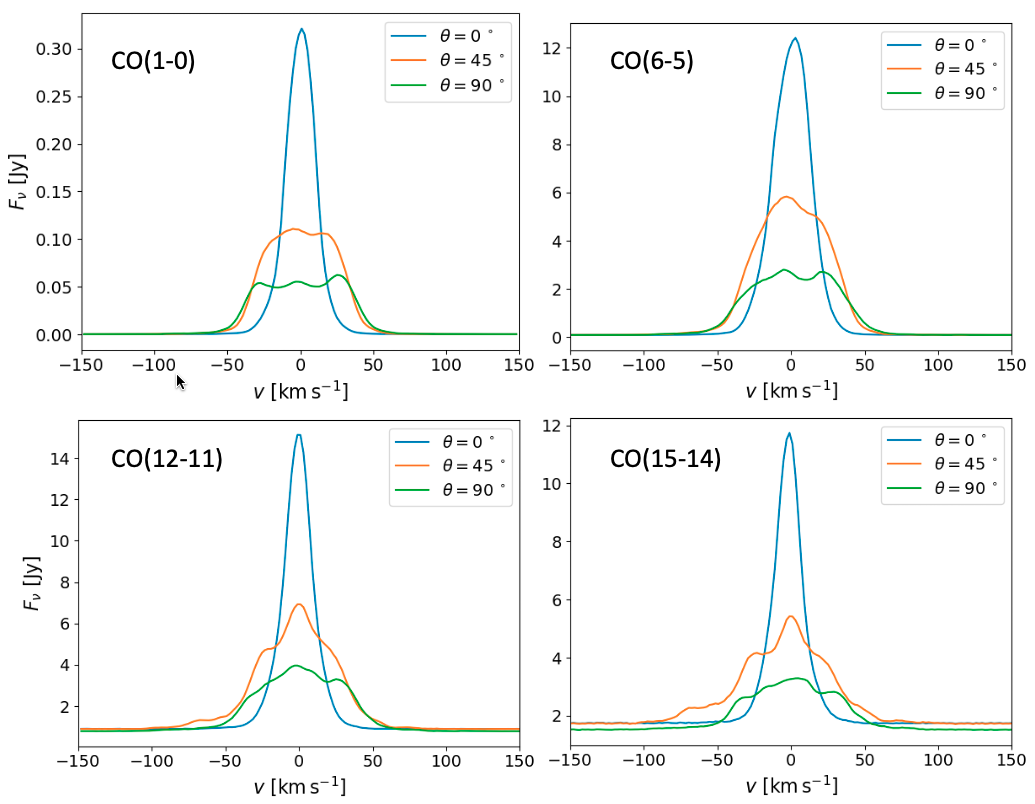}
\caption{Integrated line profile of the CO(1--0), CO(6--5), CO (12--11), and CO(15--14) transitions for the AGN torus simulation. The lines correspond to line profiles seen at different inclination angles, as the legend indicates. $v=0$ $\mathrm{km\, s^{-1}}$ corresponds the rest wavelength of each line transition. The line profiles are observed by observers located at a fiducial distance of 4.2 Mpc with a beam size of (32 pc)$^2$.}
\label{Fig: Circinus Line profiles}
\end{figure}

\subsection{Dust effects on CO lines}
\label{Application: dust extinction effect on CO lines}

To demonstrate the importance of self-consistent dust and non-LTE line radiative transfer calculations, we perform {\sc{SKIRT}} CO non-LTE radiative transfer calculations with and without dust. The presence of dust has two complementary effects on the line emission. On the one hand, it acts as a source term that affects the level populations; without dust, the CMB is the only external source term. Secondly, as shown in Fig.~{\ref{Fig: Circinus SED Lines}}, it can affect the observed CO line strengths due to attenuation. To quantify the effect of dust, we compare the CO spectral line energy distribution (SLED), that is, an intensity function of the CO rotational emission lines with the different rotational levels normalized by the intensity of CO(1--0). CO SLEDs are commonly used as a diagnostic for the excitation of CO molecules \citep[e.g.,][]{2010A&A...518L..37P, 2010A&A...518L..42V, 2011ApJ...743...94R, 2015ApJ...802...81M}.

Figure \ref{Fig: Circinus SLEDs comparison} compares the total CO SLEDs between the {\sc{SKIRT}} models without dust and with dust. We chose the edge-on orientation to maximize the potential effect of dust attenuation. We find that the values of the SLED of the model with dust are slightly lower than those of the model without dust at the higher $J$ values. The reason for this suppression of the high-$J$ line intensities in the presence of dust is that the higher-$J$ lines originate from the innermost region of the torus (see Fig.~\ref{Fig: Circinus map}i), which are surrounded by the optically thick cold dust (Fig.~\ref{Fig: Circinus map}d). 
We note that the excitation of CO molecules by the dust emission is not efficient in our simulations, although \citet{Matsumoto2022} imply that the vibrational and rotational excitation of CO molecules at higher energy states is caused by the dust continuum. This difference is due to the dust temperature near the equatorial plane. \citet{Matsumoto2022} estimated a higher dust temperature of several tens K using a simpler dust model and a different dust radiative transfer code, {\sc RADMC-3D}, while the present radiative transfer calculations set the dust temperature at around 10 K. Observations with \textit{ALMA} and \textit{JWST} will be useful to determine whether the excitation of CO molecules by dust emission is really effective. 

We conclude that self-consistent dust and non-LTE line radiative transfer calculations with {\sc SKIRT} show that higher-$J$ CO lines can be attenuated by dust, changing the shape of the SLED. For the rotational CO lines considered here, this attenuation is still modest as even the higher-$J$ rotational lines are situated in regions with modest dust optical depths. However, the same effect can be much more pronounced for atomic and molecular lines at shorter wavelengths, such as H$_2$, OH rotational, and CO rovibrational lines. We, therefore, advocate the use of self-consistent dust and non-LTE line radiative transfer such as presented in this paper to interpret observational data from \textit{ALMA}, \textit{Herschel}, and \textit{JWST}.

\begin{figure}
\includegraphics[width=0.5\textwidth]{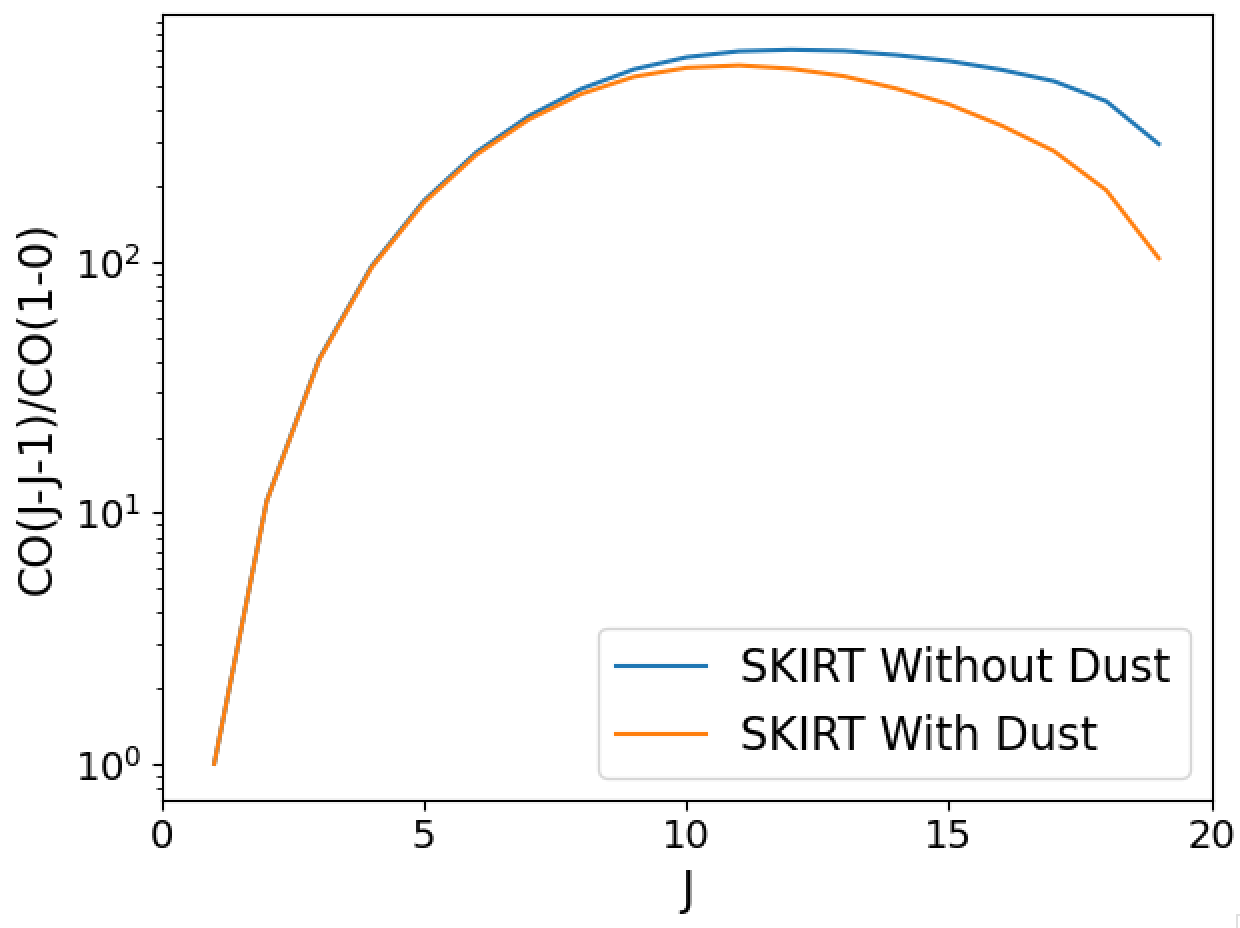}
\caption{CO SLEDs of the AGN torus model for an edge-on observer, calculated without dust (blue line) and with dust (orange line). Here, the beam size is (32 pc)$^2$.}
\label{Fig: Circinus SLEDs comparison}
\end{figure}


\section{Discussion and conclusions}
\label{Conclusion}

Many codes are available in the astronomical literature for either dust radiative transfer or non-LTE line radiative transfer, but self-consistent 3D dust and non-LTE line radiative transfer codes have not been established.
We introduce Monte Carlo-based non-LTE line radiative transfer in the 3D dust radiative transfer code {\sc SKIRT}. The result is a pure 3D Monte Carlo code that can self-consistently generate full spectra with continuum and molecular and atomic lines. The main goals of this paper are to investigate whether this approach can achieve accuracy comparable to other codes and to assess how dust affects molecular emission lines.

Our first task is to validate the accuracy of the non-LTE line radiative transfer module in {\sc SKIRT}. This has been done using the \citet{VanZadelhoff_2002A&A} benchmark models. We have compared our results with the reference results presented in this paper and the results obtained with the modern non-LTE line ray-tracing code {\sc MAGRITTE} \citep{Frederik2020}. For the spherical problem 1a/b of \citet{VanZadelhoff_2002A&A}, the level populations obtained by {\sc SKIRT} agree with those obtained by {\sc MAGRITTE} and the reference results as shown in Figure \ref{Fig: Benchmark Problem 1a/b}. For benchmark problem 2a/b, for which we use a 3D grid in {\sc{SKIRT}}, the {\sc{SKIRT}} level populations show a very satisfactory agreement with the {\sc MAGRITTE} ones with a relative difference of maximum $20\%$ (Figure~\ref{Fig: Benchmark Problem 2a/b}). This accuracy is similar to the relative accuracy reached by other codes participating in the benchmark. For problem 1b we have tested the dependence of the calculation accuracy on the number of photon packets and the optical depth of each cell. We find that the number of photon packets should be chosen such that $N_\mathrm{p} > 800 \times N_\mathrm{cell} \times N_\mathrm{line}$, and the optical depth of each cell should be chosen to be smaller than 70 to produce an accurate radial profile for the level populations. The requirement of optical depth is consistent with \citet{Camps2018}. The bottom-line is that these benchmark tests validate {\sc SKIRT} as a non-LTE line radiative transfer code.

Our second task consists of the demonstration of the new capabilities of the {\sc SKIRT} code using a 3D hydrodynamic simulation and assessing how dust affects molecular emission lines. We have performed a joint dust and non-LTE line radiative transfer simulation of the AGN torus simulation of \citet{Wada2016}. As a result, we have produced multi-band images and SEDs with CO rotational lines. We find that the low-$J$ CO emission traces the geometrically thick dusty molecular torus, whereas the higher-$J$ CO lines originate from the gas with high kinetic temperature, especially in the inner-most region (see Figure \ref{Fig: Circinus map}). The line profiles corresponding to these higher-$J$ transitions show asymmetry due to turbulent gas in the torus (see Figure \ref{Fig: Circinus Line profiles}). Furthermore, higher-$J$ CO lines are slightly attenuated by surrounding cold dust when viewed edge-on. The fact that dust extinction affects even high-$J$ rotational CO lines implies even stronger attenuation effects for atomic and molecular lines at the shorter wavelengths, such as H$_2$ lines, OH rotational lines, and CO rovibrational lines. 
In future works, we will investigate the synthetic observations for molecular lines in the MIR and NIR wavelength range. 

In addition, we note that the implementation of the non-LTE line module in {\sc{SKIRT}} makes the radiative transfer code much more powerful and versatile than `just' a non-LTE line radiative transfer code. Indeed, thanks to the modular structure of the {\sc{SKIRT}} code \citep{Camps&Baes2015, Camps2020SKIRT9}, many of the features and capabilities built into the code over the last two decades are immediately available and can be combined with the new non-LTE line capabilities without the need to reinvent the wheel. In particular,
\begin{enumerate}
\item 
The non-LTE line radiative transfer capabilities can be combined with dust radiative transfer, resulting in self-consistent calculations of dust and line emission. This is particularly useful as dust emission can promote rotational or vibrational transitions. 
We note that the most commonly used dust models in the literature \citep[e.g.,][]{Zubko2004, Draine&Li2007, Jones2017} are built into {\sc{SKIRT}}, and that the dust radiative transfer module is strongly optimized and thoroughly tested \citep{2017A&A...603A.114G}. 
\item 
{\sc{SKIRT}} is fully 3D and optimized for radiative transfer on arbitrarily complex geometries, thanks to efficient grid traversal algorithms for hierarchical \citep{Saftly2013, Saftly2014} and unstructured grids \citep{Camps2013}. The new non-LTE line radiative transfer module can immediately use these grids. As a result, {\sc{SKIRT}} is ideally suited to generate mock continuum and line observations for various types of hydrodynamical simulations, including grid-based, SPH, and moving-mesh simulations. On the other hand, the extensive library of built-in analytical geometries \citep{Baes&Camps2015} can be used to easily build phenomenological models for the interpretation of observations.
\item
The {\sc{SKIRT}} design is such that the user can easily set up multiple external radiation sources with arbitrary spectral energy distributions. Standard choices include the CMB or stellar populations with built-in or user-provided stellar population template libraries. CMB and stellar radiation can promote rotational and electronic transitions in atoms, respectively.
\item 
Non-LTE line radiative transfer calculations for different atoms and molecules (and dust) are performed simultaneously, and hence, we can consider the contamination of absorption and emission caused by other transition lines. 
Additionally, thanks to this feature, we can consider {\sc{SKIRT}} to extend to post-processing calculations for chemical reactions in future works. We note that these calculations take significant computational costs currently.
\item 
{\sc{SKIRT}} does not only perform dust and non-LTE line radiative transfer but also Ly$\alpha$ resonant line radiative transfer \citep{Camps2021} and X-ray radiative transfer \citep{Bert2023}. This offers the opportunity to explore a variety of observational tracers, covering the entire X-ray to radio regime, of the same system in a self-consistent manner. 
\end{enumerate}
The set of these features described above is unique to {\sc SKIRT}. We argue that this code is ideal for generating mock observations including both continuum and atomic and molecular lines over a wide wavelength range. The applicability is not limited to a specific class of objects but covers various astronomical fields such as galaxies, star-forming regions, and AGNs. 
Therefore, our self-consistent dust and non-LTE line radiative transfer calculations in {\sc SKIRT} can help in the interpretation of observational data from various telescopes such as \textit{Herschel}, \textit{ALMA}, and \textit{JWST}, and we invite interested colleagues to apply it to their preferred astronomical objects.

\section*{Acknowledgements}

Numerical computations were performed on the Cray XC50 at the Center for Computational Astrophysics, National Astronomical Observatory of Japan, and the {\sc SQUID} at the Cybermedia Center, Osaka University as part of the HPCI system Research Project (hp220044). 
KM gratefully acknowledges the financial support of the Flemish Fund for Scientific Research (FWO-Vlaanderen) provided through grant number 1169822N. FDC is a postdoctoral research Fellow of the Research Foundation - Flanders, grant 1253223N. This work was supported by JSPS KAKENHI Grant Number 21H04496 (KW and TN), 23H05441 (TN), 19H05810, 20H00180, and 22K21349 (KN). 

\bibliography{aanda}  
\begin{appendix} 

\section{General formalization of self-consistent non-LTE and dust radiative transfer}
\label{Ap: General formalization}

In this Appendix, we describe the formal equations that describe the self-consistent dust and non-LTE line radiative transfer calculation.
The time-independent monochromatic radiative transfer equation is expressed as
\begin{align}
\frac{\txd I}{\txd s}(\bfx,\bfn,\lambda)
=
& j(\bfx,\bfn,\lambda) 
- \alpha_{\text{abs}}(\bfx,\lambda)\,I(\bfx,\bfn,\lambda) 
\nonumber \\
& - \alpha_{\text{sca}}(\bfx,\lambda)\,I(\bfx,\bfn,\lambda) 
\nonumber \\
& + \alpha_{\text{sca}}(\bfx,\lambda) 
\int I(\bfx,\bfn',\lambda)\,\Phi(\bfx,\bfn,\bfn',\lambda)\,\txd\Omega',
\label{eq: general RT}
\end{align}
where $I$ represents the specific intensity of the radiation, which is a function of position $\bfx$, photon propagation direction $\bfn$, and wavelength $\lambda$.
$\alpha_{\text{abs}}$, $\alpha_{\text{sca}}$, and $j$ are the absorption coefficient, the scattering coefficient, and the emissivity of the medium, and $\Phi$ is the scattering phase function. 
The emissivity has contributions from the gas, the dust, and any primary radiation source,
\begin{equation}
j = j_{\text{gas}} + j_{\text{dust}}+j_\star.
\end{equation}
The absorption coefficient has contributions from the gas and dust, whereas scattering is contributed only by dust as long as we consider the radiation in the UV--submillimeter wavelength range, 
\begin{gather}
\alpha_{\text{abs}} = \alpha_{ \text{abs, gas}} + \alpha_{\text{abs, dust}},
\\
\alpha_{\text{sca}} = \alpha_{\text{sca, dust}}.
\end{gather}

Gas media such as atoms, ionized atoms, and molecules have quantized energy levels composed of electronic, vibrational, and rotational levels. The transitions between two energy levels are responsible for the line emission and absorption of the photon at a particular wavelength. As a result, the absorption coefficients and the emissivity of the gas medium are related to the probability of the transitions and are given by
\begin{gather}
  \alpha_{\text{abs, gas}} = \sum_\mathrm{line} \,
  \frac{h c}{4\pi \lambda_\mathrm{line}} \left(n_l\,B_{lu}-n_u\,B_{ul}\right) \phi_\mathrm{line}(\lambda),
  \label{eq: absorption coefficient} \\
  j_{\text{gas}} = \frac{hc}{\lambda_\mathrm{line}}\,n_u\,A_{ul}\, \phi_\mathrm{line}(\lambda).
\label{eq: emissivity} 
\end{gather}
In the equations, $A_{ul}$, $B_{ul}$, and $B_{lu}$ are the Einstein coefficients, which are related by
\begin{gather}
B_{ul} = \frac{\lambda_\mathrm{line}^5}{2 hc^2}\,A_{ul}, \\
B_{lu} = \frac{g_u}{g_l}\,B_{ul} = \frac{g_u}{g_l}\, \frac{\lambda_\mathrm{line}^5}{2 hc^2}\, A_{ul},
\end{gather}
with $g_u$ and $g_l$ the degeneracies of the upper and lower energy levels, respectively.

The absorption and scattering coefficients and the scattering phase function for the dust medium can be calculated when the sizes, chemical composition, and grain shapes are known. For spherical dust grains this can be done using Mie theory, for nonspherical grains more advanced methods need to be used \citep[e.g.,][]{1993Ap&SS.204...19V, 1994JOSAA..11.1491D, 1998JQSRT..60..309M, 2020AJ....160...55V}. The calculation of the dust emissivity depends on the absorption scattering coefficients and the temperature of the dust grains. If we assume that all dust grains are in equilibrium with the radiation field, the total emissivity of the dust medium is given by 
\begin{equation}
j_{\text{dust}} = \sum_{i=1}^{N_{\text{dust}}} \alpha_{{\text{abs}}, i} B_\lambda (T_i), 
\label{eq: dust emissivity}
\end{equation}
where the sum runs over all the different populations of dust grains, and the dust temperature $T_i$ of the $i$'th population is determined from the balance equation
\begin{equation}
\int \alpha_{{\text{abs}}, i}\, B_\lambda (T_i)\, {\text{d}}\lambda
=
\int \alpha_{{\text{abs}}, i}\, J_\lambda\, {\text{d}}\lambda.
\end{equation}
Instead of the radiative equilibrium, we can also consider stochastic heating, in which the equilibrium temperature of each grain population is replaced by a temperature probability distribution \citep[for details, see e.g.][]{Camps2015Dust}.

\section{Collisional coefficients} 
\label{Ap: Collisional coefficients}

The collisional de-excitation coefficients $C_{ul}$ depend on the number density of collisional partners $n_{\text{p}}$ and the kinetic temperature $T_{\text{kin}}$. More specifically
\begin{equation}
C_{ul} = K_{ul}(T_{\text{kin}})\, n_p,
\end{equation}
where $K_{ul}(T_\mathrm{kin})$ is the cross section of the collisional transition. The collisional de-excitation coefficients $C_{ul}$ and excitation coefficients $C_{lu}$ are related by
\begin{equation}
C_{lu} = C_{ul}\, \frac{g_u}{g_l} 
\exp\left(-\frac{h\nu}{kT_\mathrm{kin}}\right). \label{eq:equiburium_C}
\end{equation}

\end{appendix}

\end{document}